# THE CHIRAL ANOMALY, DIRAC AND WEYL SEMIMETALS, AND FORCE-FREE MAGNETIC FIELDS

**Gerald E. Marsh**

**Argonne National Laboratory (Ret)**

## ABSTRACT

The chiral anomaly is a purely quantum mechanical phenomenon that has a long history dating back to the late 1960s. Surprisingly, it has recently made a macroscopic appearance in condensed matter physics. A brief introduction to the relevant features of this anomaly is given and it is shown that its appearance in condensed matter systems must involve force-free magnetic fields, which may help explain the long current relaxation times in Dirac and Weyl semimetals.

A 2015 paper by Xiong, et al.[1] reported that the chiral anomaly, usually considered a purely quantum mechanical phenomenon, can be seen in the Dirac semimetal $Na_3Bi$. The phenomenon appears in this material when the applied electric field and magnetic fields are parallel. Because new macroscopic quantum effects are rare it is important to explore the implications of this observation.

**Some terminology and basics:**

When the mass is set equal to zero in the Dirac equation it decouples into two equations known as the Weyl equations that have two component spinors as solutions; these have charalities of $\mathcal{X} = \pm 1$. Now define the Hamiltonian of the Dirac semimetal $H(k)$ in terms of the spinor basis $\{I, \sigma_1, \sigma_2, \sigma_3\}$. If there is a $k_0$ such that the Hamiltonian satisfies $H(k_0) = 0$, in the vicinity of $k_0$ the continuity of Hamiltonian implies that it can be written as $H(k) = h(k) \cdot \sigma$. If $h_i$ are the components of $h$, the band structure of the Hamiltonian, $E_\pm(k) = \pm\sqrt{h_1^2 + h_2^2 + h_3^2}$, is called the Dirac cone, and if $h(k)$ is a linear function of $k$, the cone in $h$-space also forms a cone in $k$-space. An example of a Dirac cone is shown in the Fig. 1(a).

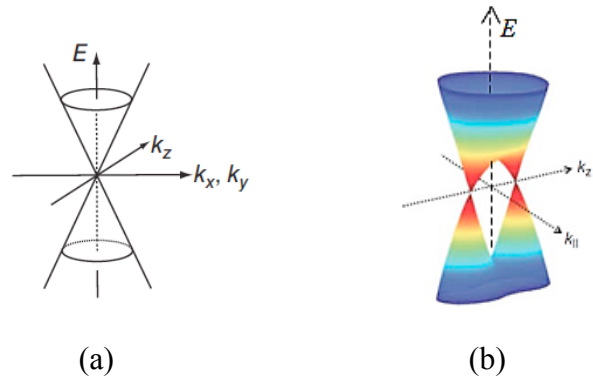

(a)  (b)

Figure 1. (a) A Dirac cone. The origin is said to have a Dirac node, and the Fermi level is located where the apexes meet. The upper cone represents the conduction band and the lower the valence band. (b) When a magnetic field is applied to a Dirac semimetal it breaks the symmetry of the crystal and causes a Dirac node to split into two chiral Weyl nodes.

The Dirac cone illustrated in Fig. 1(a) corresponds to a Dirac semimetal because there is no gap between the two cones, which would become hyperbolae when a gap is present. A normal insulator has a gap and a three dimensional topological insulator is characterized by the bulk of the material having a gap while the surface does not.

A Dirac semimetal, such as $Na_3Bi$, is a three dimensional system with a Dirac cone having a double degeneracy at the Fermi energy; a Weyl semimetal has its valence and conduction bands touching each other at isolated points, around which the band structure forms non-degenerate three dimensional Dirac cones. The apexes of the Dirac cones are called Weyl nodes. Low energy quasiparticle excitations in Weyl semimetals give the first example of the appearance of massless Weyl fermions in nature.

Figure 1(b) shows the band structure of a Dirac semimetal when a strong magnetic field is applied. If there is no electric field present, chirality is preserved at the two nodes. If, however, an electric field is applied charge will flow between the nodes, and the chiral anomaly will not vanish. The charge transfer rate depends on the chirality $\chi$ (see Eq. (3.2) below). The standard textbooks on topological insulators expand on these definitions and on the topological nature of Weyl nodes and their relation to Berry curvature.[2, 3]

The first section of this paper explains some aspects of the chiral anomaly and the second explains the connection with force-free magnetic fields and their relevance to the chiral anomaly observed in the semimetal $Na_3Bi$. The third section looks at the relaxation of such fields in a medium with a non-zero resistivity.

**1. Chiral Anomaly**

In classical physics there is said to be a symmetry when the action $S(\psi)$ is invariant under the transformation $\psi \rightarrow \psi + \delta\psi$, while in quantum mechanics the path integral $\int D\Psi\, e^{iS(\Psi)}$ must be invariant for a symmetry to be present. The transformation from classical to quantum mechanics does not always retain a given symmetry. Otherwise said: Symmetries in terms of classical, commuting variables may not be retained when expressed in terms on non-commuting quantum variables. Such a symmetry is said to have a "quantum symmetry anomaly".

The quantum symmetry anomaly of interest here is the axial anomaly, which violates the conservation of axial current. The non-conservation of chirality was discovered in the late 1960s by Adler[4] and Bell and Jackiw,[5] There is a detailed discussion of the origins of the phenomenon in the textbook by Zee,[6] and a very clear explication relevant to this work has been given by Jackiw.[7]

The axial vector current is defined as $J_5^\mu = \psi^\dagger \gamma^0 \gamma^\mu \gamma^5 \psi$. For massless fermions, $J_5^\mu$ satisfies the continuity equation $\partial_{x^\mu} J_5^\mu = 0$. Now define $P_\pm = \frac{1}{2}(I \pm \gamma^5)$ and $\psi_\pm = P_\pm \psi$ so that $\gamma^5 \psi_\pm = \pm \psi_\pm$; then if $\psi$ is a classical or quantum field operator the transformation

$$\psi \to e^{i\gamma^5 \theta}, \quad \psi_\pm \to e^{\pm i\theta} \psi_\pm$$

(1.1)

is a map between different solutions of $i\gamma^\mu \partial_{x^\mu} \psi_\pm = 0$. If one now couples this equation to an external gauge field $A_\mu$,

$$i\gamma^\mu (\partial_{x^\mu} + iA_\mu(x)) \psi(x) = 0,$$

(1.2)

then for a single Fermi field coupling to $A_\mu$ the axial vector current $J_5^\mu$ obeys the anomalous continuity equation

$$\partial_{x^\mu} J_5^\mu = \frac{1}{8\pi^2} * F^{\mu\nu}(x) F_{\mu\nu}(x),$$

(1.3)

where $*F^{\mu\nu} = \frac{1}{2}\epsilon^{\mu\nu\alpha\beta} F_{\alpha\beta}$ is the dual of the field tensor $F_{\mu\nu}(x) = \partial_{x^\mu} A_\nu(x) - \partial_{x^\nu} A_\mu(x)$. For non-Abelian fields, $A_\mu = \sum_\alpha A_\mu^\alpha T_\alpha$ and the $T_\alpha$ are anti-Hermitian matrices satisfying the Lie algebra commutators with structure constants $f_{ab}^{\ c}$; i.e., $[T_a, T_b] = \sum_c f_{ab}^{\ c} T_c$. Note that the structure constants, $f_{ab}^{\ c}$, are normalized by $tr T_a T_b = -\delta_{ab}/2$. For non-Abelian fields, Eq. (1.3) becomes

$$\partial_{x^\mu} J_5^\mu = \frac{1}{8\pi^2} tr * F^{\mu\nu}(x) F_{\mu\nu}(x).$$

(1.4)

The chiral anomaly in quantum field theory comes from two triangle Feynman diagrams associated with the decay of the $\pi^0$ particle [6].

If $A_\mu$ corresponds to the electromagnetic four potential, then Eq. (1.3) becomes

$$\partial_{x^\mu} J_5^\mu = \frac{1}{4\pi^2} \boldsymbol{E} \cdot \boldsymbol{B}.$$

(1.5)

It is this form of the anomaly that is responsible for the observations of Xiong, et al. when the electric and magnetic fields in Na$_3$Bi have collinear components. Note that the anomaly vanishes when the electric and magnetic fields are perpendicular; its non-vanishing depends on the component of $\boldsymbol{B}$ parallel to $\boldsymbol{E}$. If the medium cannot sustain a Lorentz force the fields must be either perpendicular or parallel. It is the parallel case that is of interest here. The insight that the chiral anomaly should appear in crystals is due to Nielsen and Ninomiya.[8] For a topical review of the electromagnetic response of Weyl semimetals see Burkov[9].

## 2. Chiral Anomaly and Force-Free Magnetic Fields

This section gives a short introduction to force-free magnetic fields where the current is parallel to the magnetic field, implying that the Lorentz force vanishes. In the experiment by Xiong, et al., the same condition, that the current produced by an applied electric field be parallel to the magnetic field, is also required for the non-vanishing of the chiral anomaly. The origin of the current[10] is the $\boldsymbol{E}$-field parallel to $\boldsymbol{B}$, which breaks chiral symmetry and results in an axial current.

In the Dirac semimetal Na$_3$Bi the effect of the anomaly was observed when the applied electric field and magnetic field were aligned. Xiong, et al. suggested that the large negative magnetoresistance observed implied a long relaxation time for the current. Since the non-vanishing of the anomaly depends only on $\boldsymbol{E} \parallel \boldsymbol{B}$ not vanishing, the configuration of the field responsible for the anomaly interior to the Na$_3$Bi crystal is likely to be force-free. This is because the current associated with $\boldsymbol{E}$ is parallel to $\boldsymbol{B}$, and this current is itself a source for an azimuthal magnetic field that combines with the longitudinal magnetic field applied to the Na$_3$Bi to twist the flux. It is force-free because the current associated with $\boldsymbol{E}$ is parallel to the twisted field.

It will be seen below that force-free fields have a helicity associated with them that is related to the energy stored in the field. This opens up the possibility that the decay of such fields may explain the long axial current relaxation time in Dirac and Weyl semimetals without invoking quantum mechanical processes.

Fields with $E \parallel B$ are closely related to the force-free magnetic field equations $\nabla \times B = \alpha B$ with constant $\alpha$.[11] In the experiment of Xiong, et al., the applied electric field produces a current so that, because it is only the component of the electric field parallel to the applied magnetic field that yields a non-zero chiral anomaly, this current is parallel to the applied magnetic field. This means that the field is force-free. As a consequence, since the electric field corresponds to a current, $E \parallel B$ means that $E = \beta B$, where $\beta$ is a scalar function. If $\beta$ is assumed to be a constant, Maxwell's equations can be used to show that $\beta = \pm i$, so there are no real solutions. If $\beta$ is assumed to only be a function of time, $E = \beta B$ and Maxwell's equations show that

$$\nabla \times B = \frac{\dot{\beta}}{\beta^2 + 1} B$$

(2.1)

This equation tells us is that if $E = \beta(t)B$, then $B$ must satisfy the force-free field equation. The function $\dot{\beta}(\beta^2+1)^{-1}$ in Eq. (2.1) is actually a constant, call it $\alpha$, as is shown in Appendix 1 of [11]; and this restricts the form of $\beta$ to

$$\beta = i \frac{Be^{-i\alpha t} - Ae^{i\alpha t}}{Be^{-i\alpha t} + Ae^{i\alpha t}}.$$

(2.2)

If $A$ or $B$ vanishes, $\beta = \pm i$ so that there are no real solutions; If $A = \pm B$, then $\beta = \tan \alpha t$ or $\beta = \cot \alpha t$ respectively. Equation (2.1) can then be written as

$$\nabla \times B(r) = \alpha B(r).$$

(2.3)

Thus, any magnetostatic solution to the force-free field equations can be used to construct a solution to Maxwell's equations with $E$ parallel to $B$. This is true in free space (where the solutions are standing waves, which have a vanishing Poynting vector) or when $E$ generates an electric current parallel to an external magnetic field as in the experiment of Xiong, et al.

## 3. The Chiral Anomaly and Current Relaxation Lifetime

The long axial current relaxation time in Dirac and Weyl semimetals is poorly understood and is thought to be due to near conservation of chiral charge. Burkov[12] found that there is a coupling between the chiral and total charge density, but this leads to a large negative magnetoresistance only when the chiral charge density is a nearly conserved quantity with a long relaxation time.

Consider the form of the chiral anomaly given by Eq. (1.5). Using $\boldsymbol{E} = -\partial_t \boldsymbol{A}$ and integrating over both space and time gives the helicity

$$\mathcal{H} = -\int d^3x\, (\boldsymbol{A}\cdot\boldsymbol{B}).$$

(3.1)

The integral on the right hand side is the helicity of the field, $\boldsymbol{A}\cdot\boldsymbol{B}$ being the helicity density. It plays an important role in the relaxation of magnetic fields. Because helicity is a topological invariant there are conditions under which it is conserved, but here, as will be seen below, the chiral anomaly provides a mechanism for the decay of helicity that may help explain the long current relaxation time.

Kinji, Kharzeev, and Warringa[13] have shown that a "chirality imbalance" in systems with charged chiral fermions will generate an electric current in an external magnetic field; they call this the "Chiral Magnetic Effect". Because this current also acts as a source for a magnetic field, the current flowing along the magnetic field will twist the magnetic flux and induce helicity into the field. Xiong, et al. [1] have demonstrated the converse where an applied electric current causes a charge to flow from one chiral node to another of opposite chirality. That is, application of $\boldsymbol{E}\parallel\boldsymbol{B}$ causes a charge "pumping" rate $W$ between the two chiral Weyl nodes

$$W = \chi \frac{e^3}{4\pi^2 \hbar^2} \boldsymbol{E}\cdot\boldsymbol{B},$$

(3.2)

where $\chi = \pm 1$ indicates the chirality. The "chiral imbalance" referred to above can be found by defining the number densities $\boldsymbol{n}_{L,R} = \frac{1}{2V}\int d^3x\, \psi^\dagger (1\pm\gamma^5)\psi$, where $V$ is the volume and $n_L$ corresponds to the minus sign and $n_R$ to the plus. By integrating the total axial vector current $J_5^\mu$ over space and time one can then obtain the difference in left and right chiral particles; i.e.,

$$n_L - n_R = \int d^4x\, (\partial_\mu J_5^\mu) = \frac{1}{4\pi^2} \int d^4x\, (\boldsymbol{E}\cdot\boldsymbol{B}).$$

(3.3)

The integrand of the integral on the right hand side of this equation is the chiral anomaly given by Eq. (1.5). Now differentiating Eq. (3.3) with respect to time gives

$$\frac{d}{dt}(n_L - n_R) = \frac{1}{4\pi^2} \int d^3x\, (\boldsymbol{E}\cdot\boldsymbol{B}).$$

(3.4)

Note that the integration is now over a 3-volume. If one now assumes the scalar potential vanishes and substitutes $\boldsymbol{E} = -\partial_t \boldsymbol{A}$ into Eq. (3.4), and then integrates with respect to time, $(n_L - n_R)$ may be expressed in terms of the helicity

$$n_L - n_R = -\frac{1}{4\pi^2} \int d^3x\, (\boldsymbol{A}\cdot\boldsymbol{B}) = -\frac{1}{4\pi^2}\mathcal{H}.$$

(3.5)

As a result, Eq. (3.4) can be written as

$$\frac{d}{dt}(n_L - n_R) = -\frac{1}{4\pi^2}\frac{d\mathcal{H}}{dt}.$$

(3.6)

A similar expression is readily derivable from the force-free field equation $\nabla \times \boldsymbol{B} = \alpha \boldsymbol{B}$, where $\alpha$ is again a constant. The magnetic field energy $E$ due to currents $\boldsymbol{J}$ in a volume $V$ is given by

$$E = \tfrac{1}{2} \int_V \boldsymbol{J}\cdot\boldsymbol{A}\, dV.$$

(3.7)

By taking the dot product of $\boldsymbol{A}$ with the force-free field equations and using Eq. (3.7) one obtains

$$E = \tfrac{1}{2}\alpha \int_V \boldsymbol{A}\cdot\boldsymbol{B}\, dV = \tfrac{1}{2}\alpha\mathcal{H}.$$

(3.8)

Now taking the derivative with respect to time and identifying $d\mathcal{H}/dt$ with the same quantity in Eq. (3.6) gives

$$\frac{d}{dt}(n_L - n_R) = -\frac{1}{2\pi^2 \alpha}\frac{dE}{dt}.$$

(3.9)

Thus, for force-free magnetic fields, the change in the difference of the number of left and right handed chiral particles can be related to the change in energy.

The mechanism by which the chiral anomaly allows the decay of helicity can be found by taking the time derivative of the helicity density and expanding $\partial_t(\boldsymbol{A}\cdot\boldsymbol{B})$. Using the homogeneous Maxwell equations, one can then derive the expression

$$\partial_t(\boldsymbol{A}\cdot\boldsymbol{B}) + \nabla\cdot(\Phi\boldsymbol{B} + \boldsymbol{A}\times\boldsymbol{E}) = -2\boldsymbol{E}\cdot\boldsymbol{B}.$$

(3.10)

This is a continuity equation where $\boldsymbol{A}\cdot\boldsymbol{B}$ is the helicity density, $(\Phi\boldsymbol{B} + \boldsymbol{A}\times\boldsymbol{E})$ is the helicity current (the flux of helicity), and $-2\boldsymbol{E}\cdot\boldsymbol{B}$ is a helicity "sink". The latter can be seen to make sense by writing the integral form of Eq. (3.10):

$$\partial_t\int_V \boldsymbol{A}\cdot\boldsymbol{B}\, dV + \int_V \nabla\cdot(\Phi\boldsymbol{B} + \boldsymbol{A}\times\boldsymbol{E})\, dV = -2\int_V \boldsymbol{E}\cdot\boldsymbol{B}\, dV.$$

(3.11)

The integral on the right hand side of this equation represents the resistive decay of helicity ($\boldsymbol{E} = \eta\vec{j}$ where $\eta$ is the resistivity and $\vec{j}$ is the current per unit area). The rate of relaxation is determined by $\eta$. The integrand is proportional to the chiral anomaly of Eq. (1.5).

**Summary**

After discussion of some aspects of the chiral anomaly and its form when $F_{\mu\nu}$ is the electromagnetic field tensor, it was shown that in a conducting medium such as $Na_3Bi$ when $\boldsymbol{E}\parallel\boldsymbol{B}$ the field must take the form of a force-free magnetic field. It was then shown that the current relaxation time in such media will depend on the decay of helicity, which in turn depends on the chiral anomaly and the resistivity of the medium. It is likely that this mechanism has some bearing on the long axial current relaxation time in Dirac and Weyl semimetals.